# On the thermodynamic theory of curvature-dependent surface tension

Nikolay V. Alekseechkin

Akhiezer Institute for Theoretical Physics, National Science Centre "Kharkiv Institute of Physics and Technology", Akademicheskaya Street 1, Kharkiv 61108, Ukraine
Email: n.alex@kipt.kharkov.ua

## ABSTRACT

An exact equation for determining the Tolman length (TL) as a function of radius is obtained and a computational procedure for solving it is proposed. As a result of implementing this procedure, the dependences of the TL and surface tension on radius are obtained for the drop and bubble cases and various equations of state. As one of the results of the thermodynamic study, a new equation for the dependence of surface tension on radius (curvature effect), alternative to the corresponding Tolman equation and associated with the spinodal point, is obtained. The Kelvin type equation for the equimolecular radius is shown to be exact over the entire metastability region and serves as the basis for the TL equation. The expansions of surface tension near the spinodal and binodal points show that the correction to Rusanov's linear asymptotics in the first case is a series in cubes of the radius, whereas a series in curvature holds in the second case. As a result of the analysis of these expansions, the fundamental impossibility to determine the curvature effect analytically from the binodal point is established; the computational procedure determines it from the spinodal point. It is shown that just the characteristics of the system on the spinodal, mainly the TL value at zero radius, determine the curvature effect. In general, the theory reveals a close connection between surface and bulk properties.

## I. INTRODUCTION

The dependence of surface tension on the interface radius, $\sigma(R)$, is of great importance for the nucleation theory and other phenomena related to capillarity. Ignoring the curvature effect by the classical nucleation theory[1] (CNT) leads to a discrepancy between its predictions and experimental data in the region of high supersaturations corresponding to small sizes of nuclei, where this effect just manifests itself. Tolman, in his pioneering work[2], derived a general equation for $\sigma(R)$, expressing this dependence in terms of a certain characteristic length $\delta$ called the TL now. Assuming that it is constant, $\delta = \delta_\infty$, over most of the range of sizes, he obtained the famous asymptotics $\sigma_T(R) = \sigma_\infty /(1 + 2\delta_\infty / R)$, where $\sigma_\infty = \sigma(\infty)$.

At the same time, Tolman assumed that this length may depend on $R$ in the region of small sizes. However, it turns out that just this region of the order of 1 nm corresponding to nuclei containing some tens – a few hundreds of molecules, where both $\delta$ and $\sigma$ change sharply with $R$, is at the same time the nucleation region. Simple estimates show that outside this region the nucleation work is so great that nucleation does not occur. In the nucleation region, the nucleus is so small that it cannot be considered homogeneous, as assumed by CNT, and the surface effects[3] are significant; the curvature effect is a consequence of this inhomogeneity. The development of density functional theories[4-8] was aimed to take into account this inhomogeneity and thereby to bring the theory predictions into agreement with experimental data. As additional methods, the scaling relations[9] and their subsequent development,[10,11] as well as the diffuse interface theory,[12,13] should be mentioned.

The accuracy of the above Tolman asymptotics for $\sigma(R)$ was estimated[14,15] and shown that it is valid only for large drops consisting of more than $10^6$ molecules, which is far beyond the nucleation region. Thus, although it has general scientific interest, it is useless for the nucleation theory. Nevertheless, the quantity $\delta_\infty$ providing the first step towards studying the curvature effect has been the subject of numerous studies and debates in recent decades, including the density gradient[16-19] and density functional[20-22] studies. Both negative and positive signs (depending on the research method) for this quantity were reported in literature for the drop case, starting already with Tolman's estimates giving a positive value for it.[2] At the same time, much less number of works is dedicated to determining the dependences $\delta(R)$[15,23,24] and $\sigma(R)$[24-27]. Among them, the most successful is the method of determination of both of these dependences proposed by Kashchiev[24] and based on some simple approximation.

Classical thermodynamics studies the properties of macroscopic systems (such a system is formed by a large parent phase containing a nucleus), giving various relationships between their parameters. In this way, the behavior of the thermodynamic quantity $\sigma(R)$ and, consequently, $\delta(R)$, should be determined by the macroscopic parameters of the system and its equation of state (EOS). The present study provides exact equations for these dependences within the framework of classical thermodynamics and thereby confirms this expectation. This theory also validates Kashchiev's approximation.

## II. TERMODYNAMICS OF NUCLEATION

### A. Nucleation work

We consider the formation of a spherical nucleus of a new phase within the macroscopic one-component parent phase; the latter therefore plays the role of a thermostat with temperature $T_0$ and pressure $P_0$ (in what follows, subscript 0 refers to thermostat parameters). Specifically, the following two systems are considered: a drop in a vapor and a bubble in a liquid, i.e. transitions between fluid phases of the same substance. The general equation for the work of formation of a nucleus of arbitrary size is[28,29]

$$W = \Delta E - T_0 \Delta S + P_0 \Delta V \tag{1}$$

where $\Delta E$, $\Delta S$, and $\Delta V$ are the changes in the energy, entropy, and volume of the system, respectively, when one nucleus is formed. Being a consequence of only the first and second laws of thermodynamics, this equation at the same time reflects the natural condition of nucleation – constancy of temperature and pressure of the macroscopic parent phase. This equation underlies the fluctuation theory[28] and the multivariable theory of nucleation[30-35]; a nucleus is also a [heterophase] fluctuation.

Assigning subscripts 1 and 2 to the extensive parameters of the system before and after the formation of the nucleus, respectively, as well as denoting by $\varepsilon$, $s$, and $v$ the energy, entropy and volume per one particle, respectively, we have for the energy:

$$E_1 = N_{tot}\varepsilon_0\,,\quad E_2 = E + E_0 + E_\Sigma = N\varepsilon + (N_{tot} - N)\varepsilon_0 + E_\Sigma\,,\quad N + N_0 + N_\Sigma = N_{tot} \tag{2a}$$

where the parameters without index refer to the nucleus, $N$ and $N_0$ are the numbers of particles in the nucleus and the old phase (thermostat), respectively; subscript $\Sigma$ denotes the superficial quantities for the chosen dividing surface (DS). The superficial quantities arise in the Gibbs approach[36], when a real system with a diffuse interface is compared to the reference thermodynamic system with a sharp interface, where the homogeneous macroscopic properties of both coexisting phases continue up to the DS. Thus,

$$\Delta E = E_2 - E_1 = N\Delta\varepsilon + E_\Sigma\,,\quad \Delta\varepsilon = \varepsilon - \varepsilon_0 \tag{2b}$$

Similarly, $\Delta S = N\Delta s + S_\Sigma$.

As is known, two main DS are used in the thermodynamic theory of a surface: the equimolecular (EM) DS and the surface of tension (ST)[1,36-39]. By definition, $N_\Sigma = 0$ for the EM

DS, so that $N + N_0 = N_{tot}$ in this case and $\Delta V = (V + V_0) - V_1 = (N\upsilon + N_0\upsilon_0) - N_{tot}\upsilon_0 = N\Delta\upsilon$. The following equations hold for the EM DS:[37]

$$\frac{\partial \sigma_e}{\partial T_0} = -\frac{S_\Sigma^{(e)}}{A_e}, \quad \frac{E_\Sigma^{(e)}}{A_e} = \sigma_e - T_0 \frac{\partial \sigma_e}{\partial T_0}, \quad A_e = 4\pi R_e^2 \tag{3a}$$

from where

$$E_\Sigma^{(e)} - T_0 S_\Sigma^{(e)} = \sigma_e A_e \tag{3b}$$

where $R_e$ is the EM radius of the nucleus, $\sigma_e$ is the surface tension for the EM DS. Hereafter, the quantities for the EM DS will be denoted by subscript $e$, whereas the quantities for the ST will be written without index (e. g., $R$ and $\sigma$ are the ST radius and surface tension, respectively).

Introducing the chemical potentials of a particle in the old phase, $\mu_0(T_0, P_0) = \varepsilon_0 - T_0 s_0 + P_0 \upsilon_0$, and in the nucleus, $\mu(T_0, P) = \varepsilon - T_0 s + P\upsilon$, we get

$$\Delta\varepsilon - T_0 \Delta s + P_0 \Delta\upsilon = \Delta\mu_{lv} - \Delta p\,\upsilon, \quad \Delta\mu_{lv} = \mu(T_0, P) - \mu_0(T_0, P_0), \quad \Delta p = P - P_0 \tag{4}$$

where $\Delta p$ is the Laplace pressure produced by the curved DS. For the DS of arbitrary radius $r$, it has the following form:[37]

$$\Delta p = \frac{2\sigma_r}{r} + \left[\frac{\partial \sigma_r}{\partial r}\right] \tag{5}$$

where the brackets mean that the derivative is taken with respect to the mathematical displacement of the DS.

Ono and Condo[37] obtained the following equations:

$$\sigma_r = \frac{\sigma}{3}\left[\frac{R^2}{r^2} + 2\frac{r}{R}\right] \tag{6a}$$

$$\frac{\partial \sigma_e}{\partial R_e} = \left[\frac{\partial \sigma_r}{\partial r}\right]_{r=R_e} \tag{6b}$$

The function $\sigma_r(r)$ has a minimum at the ST, i.e. $\sigma_r^{(\min)} = \sigma$ at $r = R$.

From Eqs. (5) and (6a), one obtains

$$\left[\frac{\partial \Delta p}{\partial r}\right] = 0 \tag{7a}$$

$$\Delta p = \frac{2\sigma}{R} \tag{7b}$$

i. e. the Laplace pressure as a physical quantity does not depend on the DS location $r$; it is determined by the ST, emphasizing the unique role of this DS.

As a result, Eq. (1) acquires the following form for the EM DS:

$$W_e = \frac{V_e}{\upsilon}(\Delta\mu_{\text{lv}} - \Delta p\,\upsilon) + \sigma_e A_e \tag{8}$$

where $V_e = (4\pi/3)R_e^3$ and $V_e/\upsilon = N$.

The extremum condition $\partial W_e/\partial R_e = 0$ determines the critical radius $R_e^*$ – radius of the nucleus in unstable equilibrium with the mother phase. Computing this derivative, we differentiate the quantities $V_e$, $\Delta p$, $\sigma_e$, $A_e$ and utilize Eq. (5) with $r = R_e$ together with Eq. (6b); as a result,

$$\frac{\Delta\mu_{\text{lv}}}{\upsilon} = \frac{R_e}{3}\frac{\partial\Delta p}{\partial R_e} = 0 \tag{9}$$

where the *identical* equality of the derivative to zero is obtained from Eqs. (5) and (6a) with $r = R_e$, together with Eq. (6b); it is similar to Eq. (7a) and can be viewed as its particular case:

$$\frac{\partial\Delta p}{\partial R_e} = \left[\frac{\partial\Delta p}{\partial r}\right]_{r=R_e} = 0 \tag{10}$$

Eq. (9) yields the familiar condition of phase equilibrium

$$\Delta\mu_{\text{lv}} = 0, \quad \mu(T_0, P) = \mu_0(T_0, P_0) \tag{11}$$

from which the desired critical radius $R_e^*$ must be determined; the condition $\partial W_e/\partial R_e = 0$ itself does not give this radius, as might be expected at first glance. So, an important result of the above analysis is that the equilibrium condition should give the *EM critical radius* $R_e^*$, which determines the nucleation barrier $W_e^*$; this fact will be essentially used later.

## B. Thermodynamic relations for a critical nucleus

Combining Eqs. (8) and (11), we get the work of critical nucleus formation $W_e^*$ (which is the nucleation work proper):

$$W_e = -\Delta p V_e + \sigma_e A_e \tag{12}$$

In what follows, we will deal only with critical nuclei, so critical quantities will not be marked with an asterisk.

The corresponding Gibbs equation for the work[36]

$$W = -\Delta p V + \sigma A = \frac{1}{3}\sigma A \tag{13}$$

refers to the ST; the second equality is obtained with the use of Eq. (7b).

As shown above, $(\Delta p)_e = \Delta p$, so Eqs. (12) and (13) have the same form. From Eq. (6a) with $r = R_e$, one obtains

$$\sigma_e A_e = \frac{1}{3}\sigma A + \Delta p V_e$$

and Eq. (12) gives $W_e = W$. Thus, the nucleation work has an invariant form and the same value for these DS, as it must for a physical quantity.

Differentiating Eq. (12) with respect to $\Delta p$, we use the equality $\partial/\partial\Delta p = (\partial R_e/\partial\Delta p)(\partial/\partial R_e)$ and Eq. (5) with $r = R_e$. As a result,

$$-\Delta p\frac{\partial V_e}{\partial\Delta p} + \sigma_e\frac{\partial A_e}{\partial\Delta p} + A_e\frac{\partial\sigma_e}{\partial\Delta p} = 0$$

and[24,40]

$$\frac{\partial W}{\partial\Delta p} = -V_e \tag{14}$$

The temperature $T_0$ is assumed to be constant in the present theory, so this condition is not highlighted in this and other equations and partial derivatives will sometimes be replaced by ordinary ones.

In a similar way or from Eq. (14) directly, the "conjugate" relation is derived:

$$\frac{\partial W}{\partial V_e} = -\frac{R_e}{3}\frac{\partial\Delta p}{\partial R_e} \tag{15}$$

In contrast to Eq. (9), here the derivative $\partial\Delta p/\partial R_e$ is not equal to zero; it reflects the real physical dependence $\Delta p(R_e)$ for a *critical* nucleus, like the dependence $\Delta p(R)$ considered later.

Introducing the TL[2] which is the spacing between the EM DS and ST, $\delta(R) = R_e(R) - R$, we have

$$V_e = \frac{4\pi}{3}(R+\delta(R))^3 = V + A\tilde{\delta}(R), \quad \tilde{\delta}(R) \equiv \delta(R)\left[1+\frac{\delta(R)}{R}+\frac{1}{3}\left(\frac{\delta(R)}{R}\right)^2\right] \tag{16a}$$

from where

$$\delta(R) = R\left[1+3\frac{\tilde{\delta}(R)}{R}\right]^{1/3} - R \tag{16b}$$

Another important relation involves the function $\tilde{\delta}(R)$. Differentiating Eq. (13) with respect to $\Delta p$ and taking into account the equality

$$-\Delta p\frac{\partial V}{\partial\Delta p} + \sigma\frac{\partial A}{\partial\Delta p} = 0$$

we get

$$\frac{\partial W}{\partial\Delta p} = -V + A\frac{\partial\sigma}{\partial\Delta p}$$

Comparing this equation to Eq. (14) and employing Eq. (16a), we obtain the desired relation:

$$\frac{\partial\sigma}{\partial\Delta p} = -\tilde{\delta}(R) \tag{17}$$

Eqs. (14) and (17) will be used later for surface tension expansions.

From the Gibbs adsorption equation $d\sigma = -\Gamma d\mu$, equation $d\mu = dP/\rho = dP_0/\rho_0$, and Eq. (7b), one obtains[2,36]

$$\frac{\partial\sigma}{\partial\Delta p} = -\frac{\Gamma}{\rho - \rho_0} \tag{18}$$

where $\Gamma = N_\Sigma / A$ is the adsorption for the ST; $\rho = \upsilon^{-1}$ and $\rho_0 = \upsilon_0^{-1}$ are the macroscopic densities. From Eqs. (17) and (18),

$$\tilde{\delta}(R) = \frac{\Gamma}{\rho - \rho_0} \tag{19}$$

This equation shows that the function $\tilde{\delta}(R)$ is not only a certain characteristic length, i.e. a geometrical function (as can be seen from Eq. (16a)), but also an important physical function. Both the functions, $\tilde{\delta}(R)$ and $\delta(R)$ coincide in the limit $R \to \infty$: $\tilde{\delta}(R) = \delta(R) = \delta_\infty$.

Using the definition of the adsorption $\Gamma$, Tolman[2] derived Eq. (19) by integrating the difference of densities over a transitional surface layer between coexisting phases. In the case of a drop, the lower integration limit must lie in a liquid with the properties of a macroscopic phase. However, macroscopic properties are not reached inside very small droplets, so Tolman was unsure about the correctness of Eq. (19) in this case. The present derivation avoids this problem, so Eq. (19) is true for droplets of arbitrary size; the reference thermodynamic system exists and therefore the determination of $\Gamma$ is possible for them.

### C. One more derivation of Tolman's equation and new equation for radius-dependent surface tension

Tolman[2] derived the equation for the dependence $\sigma(R)$ from Eqs. (18) and (19). On the other hand, Eq. (14) with $W = \Delta pV/2$ and Eq. (16a) or Eq. (17) alone can also serve this purpose; the latter gives the shortest way:

$$-\frac{d\sigma}{d\Delta p} = -\frac{d\sigma}{dR}\left(\frac{d\Delta p}{dR}\right)^{-1} = \frac{R}{2}\frac{\sigma'/\sigma}{R^{-1} - \sigma'/\sigma} = \tilde{\delta}(R)$$

where Eq. (7b) was used and the prime denotes the derivation with respect to $R$. From here,

$$\frac{\sigma'}{\sigma} = \frac{2\tilde{\delta}(R)/R^2}{1 + 2\tilde{\delta}(R)/R} \equiv \varphi_B(R) \tag{20a}$$

$$\sigma(R) = \sigma_\infty \exp\left[-\int_R^\infty \varphi_B(r)dr\right] \tag{20b}$$

which is the Tolman equations[2]. In this way, Tolman Eq. (20a) is derived *without using the adsorption* $\Gamma$, differently from its original derivation; the function $\tilde{\delta}(R)$ is initially defined by Eq. (16a) as merely some length, similarly to the TL $\delta(R)$.

An alternative equation for $\sigma(R)$ is derived as follows. From Eq. (7b), one obtains

$$\frac{d\ln\Delta p}{dR} = \frac{d\ln\sigma}{dR} - \frac{1}{R} = \varphi_B(R) - \frac{1}{R} = -\frac{1}{R + 2\tilde{\delta}(R)} \equiv -\varphi_S(R) \tag{21}$$

where Eq. (20a) was utilized. Integration of this equation under the physical condition that $\Delta p$ is finite on the spinodal ($R = 0$), $\Delta p(0) \equiv \Delta p_s = 2K(T_0)$, gives the desired equation:

$$\Delta p(R) = \Delta p_s \exp\left[-\int_0^R \varphi_S(r)dr\right] \tag{22a}$$

$$\sigma(R) = KR \exp\left[-\int_0^R \varphi_S(r)dr\right] \tag{22b}$$

The linear asymptotics

$$\sigma(R) = KR \tag{23}$$

in the limit $R \to 0$ is trivially obtained from Eq. (7b) and the $\Delta p$ finiteness condition in this limit. It was obtained by Rusanov[38,39] from the analysis of possible behavior of the quantity $\Gamma/(\rho - \rho_0)$ at $R \to 0$ which also uses this condition.

The difference between the two equations for $\sigma(R)$ is obvious. While Tolman's dependence $\sigma(R)$ from Eq. (20b) is "tied" to the binodal ($R \to \infty$), the dependence $\sigma(R)$ from Eq. (22b) is "tied" to the spinodal. However, this difference is deeper from the physical point of view. As mentioned above, Eq. (20b) was obtained without using adsorption; so it contains no physical characteristics of the bulk phases coexisting on the binodal. Apparently, this property relates to the fact established later as a result of a more detailed analysis that it is impossible to determine the dependence $\sigma(R)$ analytically from the binodal point. On the contrary, Eqs. (22a, b) contain an important physical quantity $\Delta p_s$ which is determined by the bulk properties of the metastable phase (by the EOS). Just these equations will play a key role in the subsequent analysis.

## III. EQUILIBRIUM CONDITION AND BASIC EQUATION FOR DETERMINING THE TOLMAN LENGTH

## A. Equilibrium condition

Writing the condition of phase equilibrium, Eq. (11), in the differential form, $d\mu_{\rm v} = \upsilon_{\rm v} dP_{\rm v} = d\mu_l = \upsilon_l dP_l$, one obtains after integration from the binodal state

$$\Delta\mu_{\rm v} = \int_{P_\infty}^{P_{\rm v}} \upsilon_{\rm v} dP_{\rm v} = \Delta\mu_l = \int_{P_\infty}^{P_l} \upsilon_l dP_l \equiv \Delta\mu \tag{24a}$$

where subscripts v and $l$ relate to vapor and liquid phases, respectively, and $\Delta\mu$ is the difference in chemical potentials between the current (supersaturated) and saturated (binodal) states. We have $P_l = P_{\rm v} \pm \Delta p$, where the plus and minus signs refer to the drop and bubble nuclei, respectively; hence,

$$\Delta\mu = \int_{P_\infty}^{P_{\rm v}} \upsilon_{\rm v} dP_{\rm v} = \int_{P_\infty}^{P_{\rm v} \pm \Delta p} \upsilon_l dP_l \tag{24b}$$

These integrals are performed using the EOS $\upsilon(P)$ for a given substance.

To get the expansion of the function $P_{\rm v}(\Delta p) - P_\infty$ in $\Delta p$ near the binodal for the drop case, we denote $\Delta p = x$, $P_{\rm v} - P_\infty = y$, and find the expansion of the function $y(x)$ from the equation

$$\int_0^y \upsilon_{\rm v}(z + P_\infty) dz = \int_0^{y+x} \upsilon_l (u + P_\infty) du \tag{25}$$

under the obvious condition $y(0) = 0$. The derivatives are as follows:

$$y'(0) = \frac{c}{1-c} \approx c, \quad y''(0) = \frac{c^2}{(1-c)^3}\left[\chi_{\rm v} - \frac{\chi_l}{c}\right] \approx c^2 \left[\chi_{\rm v} - \frac{\chi_l}{c}\right], \quad c = \frac{\upsilon_\infty^l}{\upsilon_\infty^{\rm v}},$$

$$\upsilon_\infty^l = \upsilon_l(P_\infty), \; \upsilon_\infty^{\rm v} = \upsilon_{\rm v}(P_\infty), \quad \chi_{\rm v} = -\frac{1}{\upsilon_\infty^{\rm v}} \frac{\partial \upsilon_{\rm v}}{\partial P_{\rm v}}(P_\infty), \quad \chi_l = -\frac{1}{\upsilon_\infty^l} \frac{\partial \upsilon_l}{\partial P_l}(P_\infty)$$

where $\chi_{\rm v}$ and $\chi_l$ are the isothermal compressibility coefficients of vapor and liquid, respectively, on the binodal. The condition $c \ll 1$ is assumed; e. g. $c \approx 2\times10^{-5}$ for water at room temperature and $c \approx 3\times10^{-3}$ for argon at $T_0 = 85$ °K. Thus, up to second order terms, one obtains

$$P_{\rm v}(\Delta p) - P_\infty = c\Delta p + \frac{c^2}{2}\left[\chi_{\rm v} - \frac{\chi_l}{c}\right](\Delta p)^2 \tag{26a}$$

Further, expanding the "vapor integral" $\Delta\mu_{\rm v}$ in a series in $(P_{\rm v} - P_\infty)$ up to second order terms, we have

$$\int_0^{P_\mathrm{v}-P_\infty} \upsilon_\mathrm{v}(z+P_\infty)dz = \upsilon_\infty^\mathrm{v}\left[(P_\mathrm{v}-P_\infty)-\frac{\chi_\mathrm{v}}{2}(P_\mathrm{v}-P_\infty)^2\right] = \upsilon_\infty^l\left[\Delta p-\frac{\chi_l}{2}(\Delta p)^2\right] \tag{26b}$$

where Eq. (26a) was utilized.

On the other hand, expanding the "liquid integral" $\Delta\mu_l$ in a series in $\Delta p$, we can neglect by $(P_\mathrm{v}-P_\infty)$ in comparison with $\Delta p$, in view of the condition $c \ll 1$:

$$\int_0^{(P_\mathrm{v}-P_\infty)+\Delta p} \upsilon_l(u+P_\infty)du = \upsilon_\infty^l\left[\Delta p-\frac{\chi_l}{2}(\Delta p)^2+\frac{1}{6\upsilon_\infty^l}\frac{d^2\upsilon_l}{dP_l^2}(P_\infty)(\Delta p)^3+\cdots\right] \tag{27}$$

Comparing this equation to Eq. (26b), we see that both expansions coincide, as it must. An important conclusion is these expansions are expressed only in terms of the parameters of the *liquid*, rather than vapor. The same result is obtained for the bubble case; the only difference is that the series in $\Delta p$ is alternating, where the odd terms have a minus sign. Apparently, the physical reason is that the Laplace pressure $\Delta p$ "breaks the symmetry" between the vapor and liquid phases – it acts on the *liquid* both in the case of a drop and in the case of a bubble; it compresses a liquid drop and stretches the bulk liquid phase in the bubble case. The latter can be easily seen by the example of a large negative pressure $|P_l| \gg P_\mathrm{v}$ in the liquid. The Laplace pressure $\Delta p = -P_l$ in this case ensures the mechanical equilibrium of a bubble of critical size and, consequently, stretches the liquid and tends to collapse the bubble; the coefficient $\chi_l$ is the "stretch factor" here.

The vapor spinodal pressure $P_s^\mathrm{v}$ is determined by the EOS; e.g., this is the maximum point on the van der Waals (vdW) curve $P(\upsilon)$. The corresponding $\Delta p_s^{(d)}$ value is found as a root of Eq. (24b):

$$\int_{P_\infty}^{P_s^\mathrm{v}} \upsilon_\mathrm{v}dP_\mathrm{v} = \int_{P_\infty}^{P_s^\mathrm{v}+\Delta p_s^{(d)}} \upsilon_l dP_l \tag{28a}$$

where each of the integrals represents the limiting difference of chemical potentials $\Delta\mu_s^{(d)}$ for the drop case.

The liquid spinodal pressure $P_s^l$ (the minimum point on the vdW curve) is the [theoretical] ultimate tensile strength of the liquid at a given temperature; the corresponding $\Delta p_s^{(b)}$ value is given by the equation

$$\int_{P_\infty}^{P_s^l+\Delta p_s^{(b)}} \upsilon_\mathrm{v}dP_\mathrm{v} = \int_{P_\infty}^{P_s^l} \upsilon_l dP_l \tag{28b}$$

where each of the integrals gives the limiting difference of chemical potentials $\Delta\mu_s^{(b)}$ for the bubble case. At a sufficiently low temperature, when $P_s^l$ is negative and large in absolute value, the approximation $\Delta p_s^{(b)} \approx -P_s^l$ will be used with good accuracy, as $P_{\text{v}} << \left|P_s^l\right|$ in this case.

## B. Basic equation for determining the Tolman length

As stated above, the condition of phase equilibrium, i.e. Eq. (24b), should give the EM critical radius $R_e$. On the other hand, a similar derivation of the nucleation work with the use of the ST[31] yields Gibbs Eq. (13) and the same equilibrium condition, from which the ST critical radius $R$ should be determined. Indeed, Eq. (24b) includes just the radius $R$ through $\Delta p(R) = 2\sigma(R)/R$; however, it cannot be determined from this equation until the function $\sigma(R)$ is known. So, Eq. (24b) purporting to give both radii $R_e$ and $R$, in its present form gives neither of them.

To get $R_e$ from the equilibrium condition, Eq. (24b) must be transformed to contain $R_e$ instead of $R$. Preliminarily, it should be noted that the macroscopic liquid phase in the reference thermodynamic system is under the pressure $P = P_0 + \Delta p$ in the drop case, i.e., is compressed. Thus, both the radii $R$ and $R_e$ defined using this system implicitly take into account both the compressibility and the dependence $\sigma(R)$. The ST radius $R$ in the condition of equilibrium takes into account both these phenomena by means of the liquid integral as a whole in Eq. (24b), whereas the EM radius $R_e$ in addition to $R$ contains the function $\delta(R)$ which *itself can do this* under constant values of $\upsilon_l$ and $\sigma$ in this integral. Thus, we can assume that the use of both the constant volume $\upsilon_l = \upsilon_\infty^l$ and the constant surface tension $\sigma = \sigma_\infty$ together with $R_e$ instead of $R$ in Eq. (24b) *does not change it*. Under these conditions, the liquid integral is easily performed, which yields

$$\Delta\mu = \pm\upsilon_\infty^l \frac{2\sigma_\infty}{R_e} + \upsilon_\infty^l (P_{\text{v}} - P_\infty) = \pm\upsilon_\infty^l \frac{2\sigma_\infty}{R + \delta(R)} + \upsilon_\infty^l (P_{\text{v}} - P_\infty) \quad (29)$$

The mathematical meaning of such a transformation of Eq. (24b) is as follows: the function $\delta(R)$ takes into account both the compressibility and the dependence $\sigma(R)$ through the entire metastability region, thereby giving the true dependence of $\Delta\mu$ on the degree of supersaturation; this fact gives another physical meaning to the TL. In other words, Eq. (24b) with variable quantities $\upsilon_l$, $\sigma$ and $R$ is *identical* to the same equation with constant (binodal) quantities $\upsilon_l$ and $\sigma$ when $R_e$ is used instead of $R$.

Eq. (29) gives the desired EM critical radius for a given supersaturation ratio $\xi = P_\text{v} / P_\infty$,

$$R_e(\xi) = \frac{\pm 2 v_\infty^l \sigma_\infty}{\Delta\mu(\xi) - v_\infty^l (P_\text{v} - P_\infty)} \tag{30}$$

and resembles the familiar Kelvin equation, but the latter contains $R$, not $R_e$, which is an approximation.

As a supporting physical argument in favor of the fact that the quantity $R_e$ in Eq. (29) is indeed the EM radius, we note that it has been shown both experimentally[41,42] and in density functional calculations[8,43] that Eq. (29) accurately estimates the radius $R_e$ even for very small nuclei; this fact was used in Refs. 23 and 43. Other arguments in favor of this equation will be given later, after calculations using it.

The basic equation for determining the function $\delta(R)$ is obtained from Eq. (30) as follows:

$$\delta(R) = \frac{\pm 2 v_\infty^l \sigma_\infty}{\Delta\mu(\Delta p(R)) - v_\infty^l [P_\text{v}(\Delta p(R)) - P_\infty]} - R \tag{31}$$

where either $\Delta\mu_\text{v}$ or $\Delta\mu_l$ can be taken as $\Delta\mu$. The dependence $\Delta\mu(\Delta p(R))$ is highlighted here for the procedure for determining $\delta(R)$ described later; it is given by Eq. (24b) which also defines the dependence $P_\text{v}(\Delta p)$. Interestingly, the form of this equation is consistent with the form of general Eq. (16b). Eq. (31) confirms the expectation for the function $\delta(R)$ to be determined by the macroscopic parameters of the system and its EOS.

So, both radii $R_e$ and $R$ are indeed determined by the equilibrium condition, as expected, but in quite different ways. The radius $R_e$ is determined directly and explicitly from the equilibrium condition, Eq. (30), and underlies the determination of the function $\delta(R)$. The radius $R$ is calculated only after obtaining the function $\delta(R)$ from Eq. (31) (and, consequently, the dependences $\sigma(R)$ and $\Delta p(R)$) as the root of Eq. (24b); this root $R(\xi)$ gives the desired dependence on the supersaturation. Thus, the way to get the radius $R$ is much more complicated. These facts, together with others, show the role of the EM DS in surface thermodynamics.

The spinodal value $\delta(0) \equiv \delta_s$ of the TL is

$$\delta_s^{(d)} = \frac{2 v_\infty^l \sigma_\infty}{\Delta\mu_s^{(d)} - v_\infty^l (P_s^\text{v} - P_\infty)}, \qquad \delta_s^{(b)} = \frac{-2 v_\infty^l \sigma_\infty}{\Delta\mu_s^{(b)} - v_\infty^l (P_\text{v}(\Delta p_s^{(b)}) - P_\infty)} \tag{32}$$

for drops and bubbles, respectively ($\Delta\mu_s^{(b)} < 0$).

## C. Equations of state

Quite accurate EOS for various substances are available in the literature, however, as a rule, they are quite complex and contain dozens of coefficients. To get only qualitative results illustrating the theory, some simple cubic EOS are used here. First, it is necessary to establish criteria for evaluating these EOS in relation to this theory.

The chemical potential difference for the liquid phase can be represented as follows:

$$\Delta\mu_l = \int_{\rho_\infty^l}^{\rho_l} \frac{d\rho_l}{\rho_l^2 \tilde{\chi}_l(\rho_l)} \tag{33}$$

where $\tilde{\chi}_l(\rho_l)$ is the liquid compressibility coefficient as a function of its density; $\chi_l = \tilde{\chi}_l(\rho_\infty^l)$. Thus, the compressibility coefficient $\chi_l$ and mass density $\rho_m^l$ of the liquid on the binodal obtained from the EOS will be compared with their experimental values.

On the other hand, the chemical potential difference $\Delta\mu_{\rm v}$ for the vapor phase depends on the $P_\infty$ value. The latter for a given EOS is determined by Maxwell's rule as a root of the equation

$$\int_{\upsilon_\infty^l(P_\infty)}^{\upsilon_\infty^{\rm v}(P_\infty)} P(\upsilon)d\upsilon = P_\infty\left[\upsilon_\infty^{\rm v}(P_\infty) - \upsilon_\infty^l(P_\infty)\right] \tag{34}$$

So, the triple of quantities $(\chi_l, \rho_m^l, P_\infty)$ will be used to estimate the EOS. Argon at $T_0 = 85\ K$ will be employed as a substance ($\sigma_\infty = 11.3$ dyn/cm, $T_c = 150.86\ K$, $P_c = 5\times10^7$ dyn/cm$^2$, where subscript c denotes critical point values). The vdW, Peng & Robison[44] (PR), and Guevara-Rodríguez[45] (G-R) EOS will be used for calculations.

To bring the classical vdW EOS

$$P_{vdW}(\upsilon) = \frac{kT_0}{\upsilon - b} - \frac{a}{\upsilon^2} \tag{35}$$

into accordance with the specified criteria, the coefficients $a$ and $b$ are determined from the system of equations consisting of Maxwell's rule

$$P_\infty = \frac{1}{\upsilon_\infty^{\rm v} - \upsilon_\infty^l}\left\{kT_0 \ln\frac{\upsilon_\infty^{\rm v} - b}{\upsilon_\infty^l - b} + a\left(\frac{1}{\upsilon_\infty^{\rm v}} - \frac{1}{\upsilon_\infty^l}\right)\right\} \tag{36}$$

and the $\chi_l$ definition

$$\chi_l = \frac{1}{\upsilon_\infty^l}\left[\frac{kT_0}{(\upsilon_\infty^l - b)^2} - \frac{2a}{(\upsilon_\infty^l)^3}\right]^{-1} \tag{37}$$

as the first step. The experimental values of $\upsilon_\infty^{\rm v}$, $\upsilon_\infty^l$, $P_\infty$, and $\chi_l$ are used in these equations. However, a two-parametric equation cannot keep four parameters constant. Therefore, the desired values of $P_\infty$ and $\chi_l$ are obtained from Eqs. (34) and (37), respectively, as the second step. The results presented in the Table show that this procedure gives $P_\infty$ and $\rho_m^l$ values close to the experimental ones and somewhat underestimated $\chi_l$ value. It also gives overestimated values of the critical point parameters ($T_c = 199\ K$, $P_c = 8.8\times10^7$ dyn/cm$^2$), but in the present theory this fact is considered insignificant.

The PR and G-R EOS have the following functional forms, respectively:

$$P_{PR}(\upsilon) = \frac{kT_0}{\upsilon - b} - \frac{a\alpha(T_0)}{\upsilon^2 + 2b\upsilon - b^2} \tag{38}$$

$$P_{G-R}(\upsilon) = \frac{kT_0}{\upsilon - b} - \frac{a(T_0)}{(\upsilon - c)(\upsilon - d)} \tag{39}$$

The parameters of these equations are given in original Refs. 44 and 45 in combination with the methodology for their determination; the parameters of interest calculated for these EOS are given in the Table. While the liquid density $\rho_m^l$ sometimes falls directly within the scope of the mentioned methodology (this applies to the G-R EOS), the coefficient $\chi_l$ is usually not taken into account by it. Nevertheless, the PR EOS surprisingly yields the experimental value of $\chi_l$, whereas the G-R EOS is unsatisfactory with respect to this quantity.

All three EOS under consideration are shown in Fig. 1. The vdW and G-R curves are close to each other in the liquid region, including almost the same spinodal point; the G-R curve here has the steepest slope among the three curves due to the smallest $\chi_l$ value. On the other hand, the PR and G-R curves coincide in the vapor region.

## IV. SURFACE TENSION EXPANSIONS AND COMPUTATIONAL PROCEDURE FOR DETERMINING THE TOLMAN LENGTH

### A. Surface tension expansion near the spinodal point and the possibility of its determination from this point

Eq. (14) is used to obtain the surface tension expansion near the spinodal point. Expanding the work $W$ into a series in $\Delta p$, we have

$$W(\Delta p) = -\kappa_1(\Delta p - \Delta p_s) - \kappa_2(\Delta p - \Delta p_s)^2 - \cdots \tag{40a}$$

Eq. (14) gives the expansion coefficients:

$$\kappa_1 = V_e(\Delta p_s) = V_e\big|_{R=0} \equiv V_s^e = \frac{4\pi}{3}\delta_s^3\,, \quad \kappa_k = \frac{1}{k!}\frac{d^k V_e}{d(\Delta p)^k}\bigg|_{\Delta p=\Delta p_s} \quad \text{for } k \ge 2 \tag{40b}$$

As a result of combining Eqs. (7b) and (13), the Gibbs equations[36] are obtained:

$$\sigma^3 = \frac{3}{16\pi} W(\Delta p)(\Delta p)^2 \tag{41a}$$

$$R^3 = \frac{3}{2\pi}\frac{W(\Delta p)}{\Delta p} \tag{41b}$$

Keeping only the first term in the $W(\Delta p)$ expansion and substituting it into Eq. (41a) gives

$$\sigma^3 = -\frac{3}{16\pi}(\Delta p)^2 V_s^e(\Delta p - \Delta p_s)$$

After substituting $\Delta p = 2\sigma / R$ and $\Delta p_s = 2K$, we get the first correction term to Rusanov's linear asymptotics:

$$\sigma(R) = KR\left[1 - \frac{R^3}{2\delta_s^3}\right] \tag{42}$$

By keeping the higher order terms, we can get the subsequent terms of the series in a similar way; however, this requires solving a $k$ th degree algebraic equation. To avoid this problem, we use Eq. (41b) and take into account the fact that only derivatives are needed for expansions; these derivatives can be calculated for a function given *implicitly*. Denoting

$$R^3 = y\,, \quad \Delta p = x\,, \quad \Delta p_s = x_s\,, \quad \alpha = -\frac{3}{2\pi}$$

we have for the combination of Eqs. (41b) and (40a)

$$y(x) = \alpha\frac{\kappa_1(x - x_s) + \kappa_2(x - x_s)^2 + \cdots}{x} \tag{43a}$$

The expansion

$$x(y) = x(0) + x'_y(0)\,y + \frac{1}{2}x''_{yy}(0)\,y^2 + \cdots \tag{43b}$$

where $x(0) = x_s$ and $y(x_s) = 0$, is required for our purpose.

The derivatives of the inverse function $x(y)$ are expressed in terms of the derivatives of the direct function $y(x)$ as follows:

$$x'_y = \frac{1}{y'_x}\,, \quad x''_{yy} = -\frac{y''_{xx}}{(y'_x)^3}\,, \quad x'''_{yyy} = \frac{3(y''_{xx})^2}{(y'_x)^5} - \frac{y'''_{xxx}}{(y'_x)^4}\,,\ etc. \tag{44}$$

Since $x = x_s$ corresponds to $y = 0$, we calculate the derivatives $y^{(k)}_{x^k}(x_s)$ and then find $x^{(k)}_{y^k}(0)$ using Eq. (44). In this way, the expansions of $\Delta p(R)$, Eq. (43b), and hence $\sigma(R)$ acquire the following form:

$$\Delta p(R) = \Delta p_s\left[1 + a_1 R^3 + a_2 R^6 + \cdots + a_k R^{3k} + \cdots\right] \quad (45a)$$

$$\sigma(R) = KR\left[1 + a_1 R^3 + a_2 R^6 + \cdots + a_k R^{3k} + \cdots\right] \quad (45b)$$

$$a_1 = \frac{1}{\alpha\kappa_1} = -\frac{1}{2\delta_s^3},\quad a_2 = \frac{\kappa_1 - 2K\kappa_2}{\alpha^2\kappa_1^3},\quad a_3 = \frac{\kappa_1^2 - 6K\kappa_1\kappa_2 + 4K^2(2\kappa_2^2 - \kappa_1\kappa_3)}{\alpha^3\kappa_1^5},\quad etc. \quad (45c)$$

Thus, the expansion of both the reduced Laplace pressure $\Delta p(R)/\Delta p_s$ and the function $\sigma(R)/KR$ near the spinodal point is a series in powers of $R^3$. Eq. (45b) can be viewed as an expansion for Eq. (22b).

From Eq. (20a),

$$\frac{\tilde{\delta}(R)}{R} = \frac{R\varphi_B(R)}{2(1 - R\varphi_B(R))},\quad \varphi_B(R) = \frac{d\sigma/dR}{\sigma(R)} \quad (46)$$

Substituting the $\sigma(R)$ expansion, Eq. (45b), into this equation and using Eq. (16b), we get the TL expansion as follows:

$$\delta(R) = \delta_s\left[1 + c_1 R^3 + c_2 R^6 + \cdots + c_k R^{2k} + \cdots\right] - R \quad (47a)$$

$$c_1 = \frac{2}{3}\frac{a_1^2 - a_2}{a_1},\quad c_2 = -\frac{4a_1^4 - 5a_1^2 a_2 + 9a_3 a_1 - 8a_2^2}{9a_1^2},\quad etc. \quad (47b)$$

From Eq. (47a), the linear asymptotic behavior

$$\delta(R) = \delta_s - R,\quad \delta'(0) = -1 \quad (48)$$

as the *universal* property of the TL at $R \to 0$ is obtained. It takes place for any substance at any temperature ($\delta_s$ is a function of the temperature $T_0$).

The expansion of $\Delta\mu$ in $\Delta p$ near the binodal was shown above. In a similar way, the expansion of $\Delta\mu$ in $(\Delta p - \Delta p_s)$ near the spinodal can be obtained. With the use of Eq. (45a), the general structure of the resulting expression for drops will be as follows:

$$\Delta\mu - \upsilon_\infty^l(P_\mathrm{v} - P_\infty) = \left[\Delta\mu_s - \upsilon_\infty^l(P_s^\mathrm{v} - P_\infty)\right]\left[1 + d_1 R^3 + d_2 R^6 + \cdots\right] = \frac{2\upsilon_\infty^l\sigma_\infty}{\delta_s^{(d)}}\left[1 + d_1 R^3 + d_2 R^6 + \cdots\right] \quad (49)$$

where Eq. (32) was employed. Here the coefficients $d_k$ are expressed in terms of the coefficients $a_k$ and the derivatives $d^k\upsilon_l/dP_l^k$ taken at the vapor spinodal point, i.e. at $P_l = P_s^\mathrm{v} + \Delta p_s$.

Substituting Eq. (49) into Eq. (31) and comparing the result to Eq. (47a) gives

$$\delta(R) = \frac{\delta_s^{(d)}}{\left[1 + d_1 R^3 + d_2 R^6 + \cdots\right]} - R = \delta_s^{(d)}\left[1 + c_1 R^3 + c_2 R^6 + \cdots\right] - R \quad (50)$$

Hence we conclude that (i) Eq. (31) is consistent with the expansion of $\delta(R)$ obtained from an independent thermodynamic study and (ii) the coefficients $c_k$ and, consequently, the coefficients $a_k$ (cf. Eq. (47b)) can be expressed in terms of the coefficients $d_k$, i.e. in terms of the mentioned derivatives $d^k v_l / dP_l^k$, by equating the coefficients at the same powers in the series of the LHS and RHS of Eq. (50). In this way, in principle, it is possible to determine the functions $\delta(R)$ and $\sigma(R)$ with the required accuracy. However, in practice it turns out that the series in Eqs. (45b) and (47a) converge very slowly, whereas the complexity of the coefficients $a_k$, $c_k$ and $d_k$ increases very quickly with increasing order of approximation. Therefore, this method does not allow any noticeable advance in $R$.

## B. Surface tension expansion near the binodal point and the impossibility of its determination from this point

Eq. (17) is used to obtain the surface tension expansion near the binodal point:

$$\sigma(\Delta p) = \sigma_\infty + \lambda_1 \Delta p + \lambda_2 (\Delta p)^2 + \cdots \tag{51a}$$

$$\lambda_1 = \left.\frac{d\sigma}{d\Delta p}\right|_{\Delta p=0} = -\tilde{\delta}(\infty) = -\delta_\infty, \quad \lambda_k = \frac{1}{k!}\left.\frac{d^k\sigma}{d(\Delta p)^k}\right|_{\Delta p=0} \quad \text{for } k \geq 2 \tag{51b}$$

The coefficients $\lambda_k$ relate to the function $\delta(R)$ as follows:

$$\lambda_2 = \frac{1}{2}\left.\frac{d^2\sigma}{d(\Delta p)^2}\right|_{\Delta p=0} = -\frac{1}{2}\left.\frac{d\tilde{\delta}(R)}{dR}\frac{dR}{d(\Delta p)}\right|_{R\to\infty} = \frac{1}{2}\left.\frac{d\tilde{\delta}(R)}{dR}\frac{1}{\Delta p(R)\varphi_S(R)}\right|_{R\to\infty} \tag{51c}$$

where Eq. (21) was utilized. The derivatives for $k > 2$ can be calculated by induction. The derivatives in Eq. (40b) can be transformed in a similar way.

Keeping only the linear term in this expansion,

$$\sigma = \sigma_\infty - \delta_\infty \frac{2\sigma}{R}$$

we get the Tolman asymptotics

$$\sigma_T(R) = \frac{\sigma_\infty}{1 + 2\delta_\infty / R} = \sigma_\infty \left[1 - \frac{2\delta_\infty}{R} + \frac{4\delta_\infty^2}{R^2} - \frac{8\delta_\infty^3}{R^3} + \cdots\right] \tag{52}$$

To obtain the general form of the expansion, the method described above is again applied. In the notations

$$z = \frac{1}{R}, \quad \Delta p = x, \quad \beta = \frac{1}{2}$$

Eq. (51a) takes the form

$$z(x)=\beta\frac{x}{\sigma_\infty+\lambda_1 x+\lambda_2 x^2+\cdots} \tag{53}$$

We need the expansion

$$x(z)=x(0)+x'_z(0)z+\frac{1}{2}x''_{zz}(0)z^2+\cdots \tag{54}$$

where $x(0)=0$. This is the expansion of $\Delta p$ and $\sigma$ in the *curvature* $z$:

$$\Delta p=2\sigma_\infty z\left[1+b_1 z+b_2 z^2+\cdots\right]=\frac{2\sigma_\infty}{R}\left[1-\frac{2\delta_\infty}{R}+\frac{b_2}{R^2}+\cdots\right] \tag{55a}$$

$$\sigma=\sigma_\infty\left[1+b_1 z+b_2 z^2+\cdots\right]=\sigma_\infty\left[1-\frac{2\delta_\infty}{R}+\frac{b_2}{R^2}+\cdots\right] \tag{55b}$$

Eq. (55b) can be viewed as an expansion for Eq. (20b) which has the following form in terms of the curvature:

$$\sigma(z)=\sigma_\infty\exp\left[-\int_0^z\frac{2\tilde{\delta}(z')dz'}{1+2z'\tilde{\delta}(z')}\right] \tag{56}$$

The coefficients $b_k$ are determined with aid of Eq. (44):

$$b_1=2\lambda_1=-2\delta_\infty,\ \ b_2=4\left[\lambda_1^2+\sigma_\infty\lambda_2\right],\ \ b_3=8\left[\lambda_1^3+3\sigma_\infty\lambda_1\lambda_2+\sigma_\infty^2\lambda_3\right],\ \ etc. \tag{57}$$

Using these coefficients and comparing Eq. (55b) to Eq. (52), we obtain the difference between the true function $\sigma(R)$ and Tolman's approximation:

$$\sigma(R)-\sigma_T(R)=\sigma_\infty\left[\frac{4\sigma_\infty\lambda_2}{R^2}+\frac{8\left(\sigma_\infty^2\lambda_3-3\sigma_\infty\delta_\infty\lambda_2\right)}{R^3}+\cdots\right] \tag{58}$$

Noting additionally that $b_4=16\lambda_1^4+\cdots=16\delta_\infty^4+\cdots$, we can assume by induction that the expansion for $\sigma(R)$ includes Tolman's asymptotics *as a component*, making a correction to it; this correction starts with a quadratic term. The coefficient $b_2$ is known in the literature as the rigidity constant;[17,18,22] it can be evaluated with the aid of Eq. (51c) after obtaining the function $\delta(R)$.

Further, the TL expansion is obtained, as before; preliminarily, all the necessary equations, including Eq. (16b), are converted from $R$ - to $z$ -dependence:

$$\delta(z)=c_1+c_2 z+c_3 z^2+\cdots=\delta_\infty+\frac{c_2}{R}+\frac{c_3}{R^2}+\cdots \tag{59a}$$

$$c_1=-\frac{b_1}{2}=\delta_\infty,\ \ c_2=\frac{3}{4}b_1^2-b_2,\ \ c_3=-\frac{29}{24}b_1^3+\frac{5}{2}b_1 b_2-\frac{3}{2}b_3,\ \ etc. \tag{59b}$$

The rigidity constant is determined by the derivative of the function $\delta(R)$, i.e. by the coefficient $c_2$ ($\varphi_S(R)\to 1/R$ and $\Delta p(R)=2\sigma_\infty/R+O(1/R^2)$ at $R\to\infty$ ).

Eq. (30) converted from $R$ to $z$ reads:

$$1+z\delta(z)=\pm\frac{2v_\infty^l\sigma_\infty z}{\Delta\mu-v_\infty^l(P_\mathrm{v}-P_\infty)} \tag{60}$$

Expanding the denominator according to Eq. (27) and utilizing Eq. (55a), we have

$$1+z\delta(z)=\frac{\pm 2v_\infty^l\sigma_\infty z}{\pm 2v_\infty^l\sigma_\infty z[1+b_1z+\cdots][1\mp(\chi_l/2)(2\sigma_\infty z+\cdots)]}$$

from where

$$1+z\delta(z)=1-(b_1\mp\chi_l\sigma_\infty)z+O(z^2)\,,\quad \delta(z)=-(b_1\mp\chi_l\sigma_\infty)+O(z) \tag{61}$$

Thus, $\delta_\infty=-(b_1\mp\chi_l\sigma_\infty)$. Substituting $b_1=-2\delta_\infty$, we get $\delta_\infty$ for drops[23] and bubbles:

$$\delta_\infty^{(d)}=-\chi_l\sigma_\infty\,,\quad \delta_\infty^{(b)}=\chi_l\sigma_\infty=-\delta_\infty^{(d)} \tag{62a}$$

If the correction $P_\mathrm{v}(\Delta p)-P_\infty=\pm c\Delta p$ is taken into account in Eq. (27), the $\delta_\infty$ values receive the corresponding correction:

$$\delta_\infty^{(b,d)}=\pm\chi_l\sigma_\infty(1+2c) \tag{62b}$$

The quantity $\chi_l\sigma_\infty$ was called "the fundamental length characteristic of liquids" in Ref. 46; it varies in the narrow range $0.017\div 0.047$ *nm*, i.e. by a factor less than 3, for 30 liquids of very different nature considered therein near the triple point, while $\chi_l$ and $\sigma_\infty$ separately vary by a factor of 150. In this way, the TL limiting value $\delta_\infty$ turns out to be equal in absolute value to this important characteristic of liquids. The reason why $\delta_\infty$ contains the compressibility of a liquid rather than a vapor was explained above. Eq. (62a) reflects the property of the function $\delta(R)$ mentioned above to take into account both the compressibility of the liquid and the dependence $\sigma(R)$ in Eq. (29); we can say that this property manifests itself on the binodal through Eq. (62a).

Returning to Eq. (60) for drops,

$$\Delta\mu-v_\infty^l(P_\mathrm{v}-P_\infty)=\frac{2v_\infty^l\sigma_\infty z}{1+z\delta(z)} \tag{63}$$

and expanding both sides in $z$ (Eq. (27) together with Eq. (55a) is employed for the LHS), we obtain $f_{LHS}(z)=f_{RHS}(z)$ after reducing both sides by the factor $(2v_\infty^l\sigma_\infty)$, where

$$f_{LHS}(z)=z\{1+(b_1+2\sigma_\infty d_1)z+(b_2+4b_1d_1\sigma_\infty+4d_2\sigma_\infty^2)z^2$$

$$+[b_3+4b_2d_1\sigma_\infty+b_1(2b_1d_1\sigma_\infty+12d_2\sigma_\infty^2)+8d_3\sigma_\infty^3]z^3+\cdots\},\quad d_k=\frac{1}{v_\infty^l}\frac{d^kv_l}{(k+1)!dP_l^k}(P_\infty) \tag{64a}$$

$$f_{RHS}(z)=z\left\{1+\frac{b_1}{2}z+\left(b_2-\frac{b_1^2}{2}\right)z^2+\left(\frac{3b_3}{2}-\frac{3b_1b_2}{2}+\frac{7b_1^3}{12}\right)z^3+\cdots\right\} \tag{64b}$$

In particular, $d_1 = -\chi_l / 2$.

In view of Eq. (62a), we see that the linear terms in braces coincide for these functions. However, an attempt to equate the coefficients at other powers of $z$ fails: when the coefficients at $z^2$ are equated, the coefficient $b_2$ drops out of this equation and therefore is not determined; coefficients at other powers of $z$ contain $b_2$. Thus, the coefficients at $z^k$ in the functions $f_{LHS}(z)$ and $f_{RHS}(z)$ are not equal to each other for $k \geq 2$. This means that here the coefficients $b_k$ cannot be determined in terms of the coefficients $d_k$ (in contrast to the case of expansion near the spinodal) and therefore the functions $\sigma(z)$ and $\delta(z)$ are *not determined from the binodal point* analytically (of course, these coefficients can be determined in another way, e.g., by fitting to the solution of exact Eq. (31) or to the results of density functional calculations). The same conclusion holds for the bubble case.

Thus, the functions $f_{LHS}(z)$ and $f_{RHS}(z)$ should be equal to each other *as a whole*, without the equality of the corresponding terms of their series. In other words, the series in Eqs. (64a, b) should coincide *asymptotically*: for a given $z$, the difference between them should become arbitrary small by taking a sufficient number of their terms; the higher the order of approximation, the closer these series are to each other.

Employing the expansion in Eq. (27) to the denominator of Eq. (30), we have, up to a quadratic term

$$R_e(\Delta p) = \frac{\pm 2 v_\infty^l \sigma_\infty}{\pm v_\infty^l \Delta p [1 \mp (\chi_l / 2) \Delta p]} = \frac{2\sigma_\infty}{\Delta p}\left[1 \pm \frac{\chi_l}{2}\Delta p\right]$$

which gives, in view of Eq. (62a),

$$R_e^{(K)}(\Delta p) = \frac{2\sigma_\infty}{\Delta p} - \delta_\infty^{(d,b)} \tag{65}$$

This simple approximation for $R_e(\Delta p)$, obtained in a different way, was used by Kashchiev[24] in his approach to calculating the dependences $\delta(R)$, $\sigma(R)$, and other functions of interest. The accuracy of this approximation will be verified later, after the function $\delta(R)$ has been calculated from exact Eq. (31). Since this equation is derived from an expansion near the binodal, the maximum deviation is expected to be near the spinodal. This approximation corresponds to the linear term in the braces of Eqs. (64a, b), and $f_{LHS}(z) = f_{RHS}(z) = z(1 - \delta_\infty z)$ in this case.

From Eq. (65), the function $\delta(R)$ is also determined as

$$\delta^{(K)}(R) = \frac{2\sigma_\infty}{\Delta p(R)} - \delta_\infty - R \tag{66}$$

and the following simple relation between the main parameters of the theory is obtained:

$$\delta_s^{(K)} + \delta_\infty = \frac{\sigma_\infty}{K} \tag{67}$$

where two parameters relate to the spinodal and two to the binodal. This equation explicitly shows that the $\delta_s$ value is determined mainly by the limiting Laplace pressure $\Delta p_s$ ($\delta_s$ is noticeably large than $\delta_\infty$).

Combining Eqs. (22a) and (66), we get a more explicit integral equation for $\delta^{(K)}(R)$:

$$\delta^{(K)}(R) = \frac{\sigma_\infty}{K} \exp\left[\int_0^R \frac{dr}{r + 2\tilde{\delta}^{(K)}(r)}\right] - \delta_\infty - R \tag{68}$$

where

$$\tilde{\delta}^{(K)}(R) \equiv \delta^{(K)}(R)\left[1 + \frac{\delta^{(K)}(R)}{R} + \frac{1}{3}\left(\frac{\delta^{(K)}(R)}{R}\right)^2\right]$$

## C. Computational procedure for determining the dependences $\delta(R)$ and $\sigma(R)$

As shown above, the functions $\delta(R)$ and $\sigma(R)$ cannot be determined by their expansions near the spinodal and binodal points due to technical complexity and fundamental impossibility, respectively. Therefore, these functions will be determined here *as a whole* in some range in $R$ with the aid of Eqs. (22a), (24b) and (31). The function $\Delta p(R)$ in Eq. (22a) contains the function $\delta(R)$ in the integrand; the function $\Delta\mu(\Delta p)$ in Eq. (24b) contains this $\Delta p$ in the integration limit. Thus, Eq. (31) is a "super-integral" equation for $\delta(R)$**.** It can be solved by the method of successive approximations. The computational procedure includes the following steps.

(i) The linear function $\delta_0(R) = \delta_s - R$, Eq. (48), is taken as the initial (zero) approximation for $\delta(R)$. Eq. (22a) in this case gives

$$\Delta p_0(R) = \Delta p_s \frac{2\delta_s^3}{R^3 + 2\delta_s^3} \tag{69}$$

where any value of $\delta_s$ can be taken; its true value, Eq. (32), will be determined in step (v).

(ii) The selected range in $R$, $[0, R_{\max}]$, is divided by $N_p$ points $R_i$.

(iii) The function $\Delta p_0(R_i)$ is calculated at these points.

(iv) The function $\Delta\mu(\Delta p_0(R_i))$ is calculated using the dependence $P_v(\Delta p)$ determined preliminarily, Eq. (24b).

(v) The function $\delta_1(R_i)$ is calculated at the points $R_i$ according to Eq. (31).

(vi) Cubic spline interpolation is applied to obtain the fist approximation function $\delta_1(R)$ in the selected interval $[0, R_{\max}]$.

(vii) The first approximation function $\Delta p_1(R)$ is determined by Eq. (22a) with $\delta(R) = \delta_1(R)$; hence the corresponding approximation for the surface tension $\sigma_1(R) = \Delta p_1(R)R/2$ is obtained, Eq. (22b) .

This procedure is then repeated from step (iii): the functions $\Delta p_1(R_i)$ and $\Delta\mu(\Delta p_1(R_i))$ are calculated to obtain the second approximation function $\delta_2(R)$; it determines the second approximation functions $\Delta p_2(R)$ and $\sigma_2(R)$. In this way, repeating this procedure the necessary number of times, we will determine the desired functions $\delta(R)$ and $\sigma(R)$ with a required accuracy in the given interval of $R$ values.

A similar procedure can be applied to Eq. (68); it is much simpler here, as the functions $P_v(\Delta p)$ and $\Delta\mu(\Delta p)$ are not needed. Instead, this equation uses only three parameters $\Delta p_s = 2K$, $\sigma_\infty$ ,and $\chi_l$ taken from the experiment ($\sigma_\infty$, $\chi_l$) and the EOS ($\Delta p_s$, $\chi_l$). The drop and bubble cases differ only in the value of $\Delta p_s$ and the sign of $\delta_\infty$.

## V. RESULTS AND DISCUSSION

Fig. 2 shows the result of the computational procedure. It is clear that the approximations $\delta_k(R)$ converge to a solution; as they oscillate, the average $(\delta_k + \delta_{k+1})/2$ of the odd end even approximations can be taken as an approximation for the TL in the selected interval of radii. It can be seen that for $\sigma(R)$ the convergence to a solution occurs faster than for $\delta(R)$. It should also be noted that the convergence "accelerates" as the number $k$ increases: while thirty iterations for $\delta(R)$ are enough to cover a range of about 10 nm in $R$, fifty iterations already cover a range of 200 nm. A similar picture is obtained, when this computational procedure is applied to Eq. (68).

It is worth noting that the function $\delta(R)$ determined from the spinodal point tends to its asymptotic value $\delta_\infty$ determined from expansions near the binodal. Since the PR EOS gives practically the experimental value of $\chi_l$, the corresponding value of $\delta_\infty$= -0.023 nm can be assumed as a true value for argon at this temperature. So, the function $\delta(R)$ is completely determined from the spinodal point.

It turns out that the attempt to perform this computational procedure from the binodal point fails for Eqs. (31) and (66) converted to the curvature $z$ and using Tolman's Eq. (56) for $\Delta p$. This result reflects the fact of fundamental impossibility to determine analytically the function $\delta(R)$ from this point which was established above and can be explained by the example of Eq. (66). Being converted to $z$ and using Eq. (56), this equation *does not contain* $\Delta p_s$, whereas just this quantity determines both the slope $K$ of the dependence $\sigma(R)$ at $R \to 0$ and the $\delta_s$ value, Eq. (67); i. e., this equation contains no information about the behavior of the functions $\delta(R)$ and $\sigma(R)$ near the spinodal. Thus, somewhat paradoxical situation takes place for Eq. (66): although it was obtained from expansions near the binodal, nevertheless, it is solved from the spinodal point; therefore, its accuracy near this point is important.

An additional reason is as follows. The line $P = P_\infty$ in Fig.1 separates two metastable states: $P > P_\infty$ (metastable vapor, the drop case) and $P < P_\infty$ (metastable liquid, the bubble case), each with *its own* dependence $\sigma(R)$. Therefore, the binodal state, as a borderline state between these two, "contains no information" about this dependence; the function $\delta(R)$ "cannot make a choice" between its two possibilities $\pm|\delta_\infty|$. On the other hand, each of these metastable states has its own spinodal point which gives rise to determining the mentioned dependences.

The fact that the dependence $\sigma(R)$ asymptotically tends to $\sigma_\infty$ is a strong argument in favor of the correctness of Eq. (31). In this regard, the importance of the $\upsilon_\infty^l(P_\mathrm{v} - P_\infty)$ term in the denominator of this equation should be noted; without this term, slightly underestimated values of $\sigma_\infty$ are obtained in the drop case (i. e., the $\sigma(R)$ curve crosses the line $\sigma = \sigma_\infty$). The same computational procedure applied to Eq. (68) gives the dependence $\sigma(R)$ asymptotically tending to $\sigma_\infty$ with the same accuracy, so Eq. (68) retains this important property.

The dependences $\delta(R)$ and $\sigma(R)$ for drops and various EOS are shown in Fig. 3. It is interesting that these dependences are close for the PR and vdW EOS, whereas other pairs of the EOS curves, (G-R, vdW) and (G-R, PR), are close in the liquid and vapor regions, respectively (Fig. 1). This fact is explained by the closeness of the $\Delta p_s$ values for these EOS, which leads to the closeness of the $\delta_s$ values given in the Table. In turn, close $\delta_s$ values lead to close dependences $\sigma(R)$.[47]

The dependences $\delta(R)$ and $\sigma(R)$ for bubbles are presented in Fig. 4. Here, the curves for the vdW and G-R EOS are close to each other, in accordance with the fact that the corresponding EOS curves in Fig. 1 have practically the same spinodal point $(\upsilon_s^l, P_s^l)$ in the liquid region. Therefore, the values of $\Delta p_s$ and $\delta_s$ given in the Table are close for these EOS; as mentioned above, $\Delta p_s^{(b)} = -P_s^l$.

Fig. 5 gives a comparison of the exact, $\delta(R)$, and approximate, $\delta^{(K)}(R)$, functions, as well as the corresponding $\sigma(R)$ dependences. As can be seen, these functions are close, but cross. Their maximum relative difference (at the spinodal point) is about 1.3 and 2.5 % for drops and bubbles, respectively, in the case of the PR EOS. For the other two EOS, this difference is smaller; the reason for their "higher accuracy" for Eq. (68) is explained below.

Fig. 6a gives a comparison of the exact and approximate, $R_e^{(K)}(\Delta p)$, dependences $R_e(\Delta p)$. As expected, the maximum error is at the spinodal point; the relative errors $\varepsilon_R = [R_e^{(K)} - R_e]/R_e$ here are equal to 1.8 and 2.6 % for drops and bubbles, respectively, and the PR EOS. Since Eq. (65) is based on the replacement of the denominator $\Delta\mu_-(\Delta p) = \Delta\mu(\Delta p) - \upsilon_\infty^l [P_\mathrm{v}(\Delta p) - P_\infty]$ with the parabolic approximation $\Delta\mu_{ap}(\Delta p) = \pm\upsilon_\infty^l \Delta p[1 \mp (\chi_l/2)\Delta p]$, Fig. 6b shows both these functions together with the dependence $\Delta\mu(\Delta p)$, all in $kT_0$ units. The relative error $\varepsilon_\mu = (\Delta\mu_- - \Delta\mu_{ap})/\Delta\mu_-$ is consistent with $\varepsilon_R$ in Fig. 6a.

Summarizing, we can conclude that the approximations given by Eqs. (65) and (66) are sufficiently accurate, so they can be used in practice due to their simplicity. The reason is the *smallness of the compressibility factor* $\chi_l$ for liquids, which is the curvature of the approximating parabola. For this reason, the vdW and especially G-R EOS having less values of $\chi_l$ give higher accuracy to these approximations (for $\delta_s$, this fact can be seen from the Table data). The derivatives of compressibility with respect to pressure are also small, so the liquid branch of the EOS is close to a straight line; this fact justifies the sufficiency of the mentioned parabolic approximation. In the bubble case, some deviation from the straight line occurs in the vicinity of the spinodal point, which gives somewhat higher errors for the dependences of interest.

As can be seen from Fig. 6b, the dependence $\Delta\mu(\Delta p)$ is close to the straight line $\Delta\mu_{str}(\Delta p) = (\Delta\mu_s/\Delta p_s)\Delta p$; the maximum relative deviations are equal to 3.3 and 6.3 % for drops and bubbles, respectively, and the PR EOS. This means that the curves $\Delta\mu(R)$ and $\Delta p(R)$ are geometrically similar, so the corresponding reduced dependences $\Delta\mu(R)/\Delta\mu_s$ and $\Delta p(R)/\Delta p_s$ almost coincide. Fig. 7 confirms this conclusion and shows that both $\Delta\mu$ and $\Delta p$ can be equally used as a measure of metastability. For bubbles the coincidence is somewhat worse than for drops due to the fact that the dependence $\Delta\mu(\Delta p)$ for bubbles in Fig. 6b is "more parabolic" than for drops.

Fig. 8 shows the function $\varphi_S(R)$ together with its asymptotics

$$\varphi_S(R) = \begin{cases} \varphi_S^{(0)}(R) = 3R^2/(R^3 + 2\delta_s^3), & R \to 0 \\ \varphi_S^{(\infty)}(R) = 1/R, & R \to \infty \end{cases} \tag{70}$$

where the equation for $\varphi_S^{(0)}(R)$ is obtained using the linear asymptotics from Eq. (48). The asymptotics $1/R$ reduces the factor $R$ in Eq. (22b), providing the condition $\sigma(R) \to \sigma_\infty$. Thus, the sharp dependence $\sigma(R)$ is determined by a "bell" near the spinodal point. From Eqs. (22b), (67), and the condition $\sigma(R)/\sigma_\infty \to 1$, one obtains

$$\lim_{R\to\infty} R \exp\left[-\int_0^R \frac{dr}{r + 2\tilde{\delta}^{(K)}(r)}\right] = \delta_s^{(K)} + \delta_\infty \tag{71}$$

regardless of the *shape* of the $\delta^{(K)}(R)$ curve; only its *endpoints* matter. In view of the fact that $\delta_s$ is noticeably larger than $\delta_\infty$ (cf. the Table), these results together with Figs. 3 and 4 show that the curvature effect is determined mainly by the properties of the system at the spinodal point, which confirms the earlier conjecture.[47] This equation explains why the dependences $\sigma(R)$ for functions $\delta(R)$ with the same value of $\delta_s$ are close to each other; the slight difference between them is associated only with the small quantity $\delta_\infty$.

## VI. CONCLUSIONS

The main result of the theory is Eqs. (31), (22a, b) and the corresponding computational procedure for determining $\delta(R)$ in coupling with $\sigma(R)$. The underlying thermodynamic consideration includes the following results. (i) The work of formation of a nucleus of arbitrary size derived for the EM DS, Eq. (8), and then the extremum condition applied to it lead to the conclusion that the equilibrium condition should give the EM critical radius. (ii) This conclusion, together with the nucleation work for the EM DS, Eq. (12), as well as Eqs. (14) and (17) employed for surface tension expansions, shows that the EM DS is not merely some kind of auxiliary DS, but is inherent in thermodynamics alongside with the ST; one can say that it plays the role of a kind of "zero point" in surface thermodynamics in a broad sense, not only in relation to adsorption. (iii) New Eqs. (22a, b) associated with the spinodal point are key to determining both the $\delta(R)$ and $\sigma(R)$ dependences.

Analysis of the equilibrium condition expressed by Eq. (24b) results in the conclusion that Kelvin type Eq. (29) for the EM radius is exact, which implies the physical meaning of the TL as a function that takes into account both the compressibility of the liquid and the dependence $\sigma(R)$. This equation yields basic Eq. (31) for determining the TL.

The expansion of surface tension into a series near the spinodal and binodal points shows the fundamental possibility and impossibility, respectively, of determining it from these points. The physical reason for this impossibility can be explained by the fact that Tolman's Eq. (20b) associated with the binodal contains a lack of information about the properties of the system. These expansions also provide some fundamental properties of the TL and useful equations: (i) the linear asymptotic behavior of the TL at the spinodal point, Eq. (48), as its universal property; (ii) Eq. (62a) for the asymptotic value $\delta_\infty$ of the TL at the binodal point, and (iii) approximate Eq. (65) for $R_e(\Delta p)$ and the simplified Eq. (68) for determining the TL.

The solution of Eq. (31) obtained by the method of successive approximations shows that the TL strongly depends on the radius in the region of small sizes, in accordance with Tolman's original assumption and more recent studies. This fact, together with Eq. (71) and Fig. 8, leads to the conclusion that the curvature effect is determined by the properties of the system at the spinodal point, mainly by the value of $\delta_s$. An assessment of the accuracy of approximate Eq. (68) for $\delta(R)$ shows its suitability for practical use; the reason of its good accuracy is established to be the smallness of the liquid compressibility factor $\chi_l$.

## ACKNOWLEDGMENT

I am very grateful to Dr. R. I. Kholodov from the Sumy Institute of Applied Physics, who supported me from the first tragic days of the war and kindly provided me with the conditions to carry out this work.

## AUTHOR DECLARATIONS

### Conflict of Interest

There are no conflicts of interest to declare

## DATA AVAILABILITY

The data that support the findings of this study are available within the article.

| | $\chi_l \times 10^{10}$ cm²/dyn | $P_\infty \times 10^{-5}$ dyn/cm² | $\rho_m^l$ g/cm³ | $P_s^v \times 10^{-6}$ dyn/cm² | $\Delta p_s^{(d)} \times 10^{-8}$ dyn/cm² | $\delta_s^{(d)} / \delta_s^{(K,d)}$ nm | $\delta_\infty^{(d)}$ nm | $\Delta p_s^{(b)} \times 10^{-8}$ dyn/cm² | $\delta_s^{(b)} / \delta_s^{(K,b)}$ nm |
|---|---|---|---|---|---|---|---|---|---|
| vdW | 1.23 | 7.8 | 1.45 | 11.06 | 6.13 | 0.3794/ 0.383 | -0.014 | 4.63 | 0.464/ 0.474 |
| PR | 2.05 | 8.1 | 1.65 | 8.78 | 6.23 | 0.3791/ 0.386 | -0.023 | 3.46 | 0.615/ 0.631 |
| G-R | 0.76 | 7.78 | 1.36 | 8.77 | 5.07 | 0.452/ 0.454 | -0.0086 | 4.67 | 0.469/ 0.476 |
| Experimental | 2.02 | 7.9 | 1.41 | | | | | | |

Table. Main parameters of the theory for argon at $T_0 = 85\ K$ for the EOS under consideration.

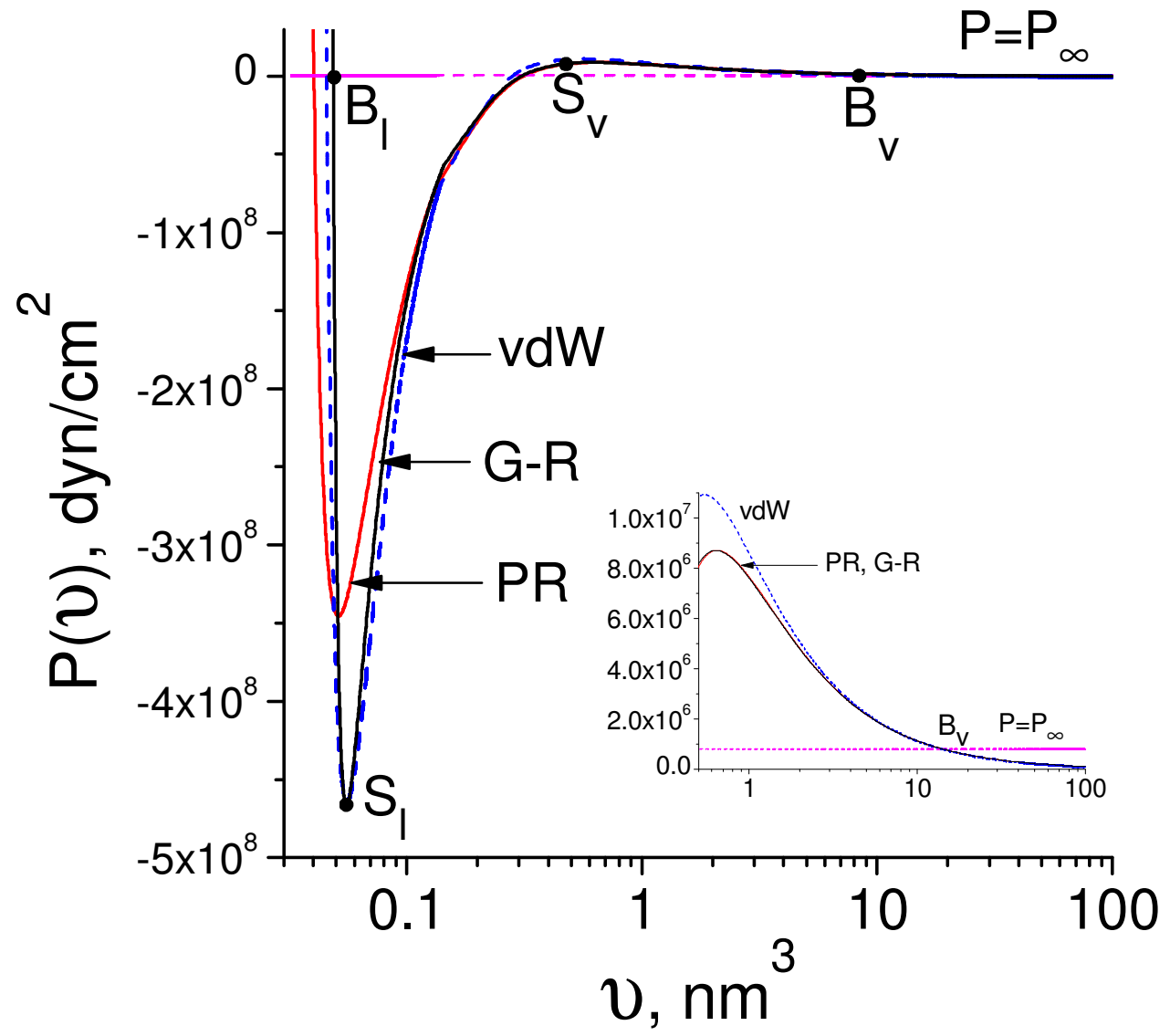

Fig. 1. EOS under consideration for argon at $T_0 = 85\ K$. $(B_v, B_l)$ and $(S_v, S_l)$ are the binodal and spinodal points for vapor and liquid, respectively.

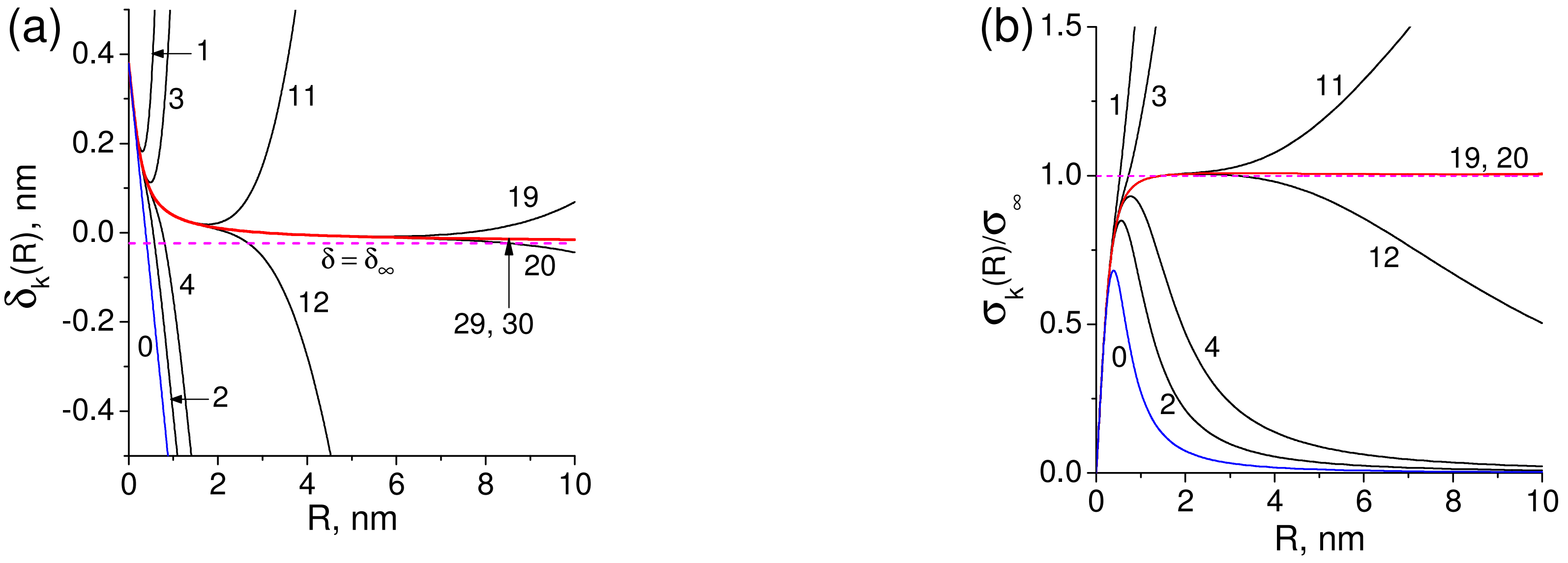

Fig. 2. Approximations $\delta_k(R)$ for the TL (a) and $\sigma_k(R)$ for the surface tension (b) in the case of drops and the PR EOS . The number $k$ of iteration is shown at the corresponding curve.

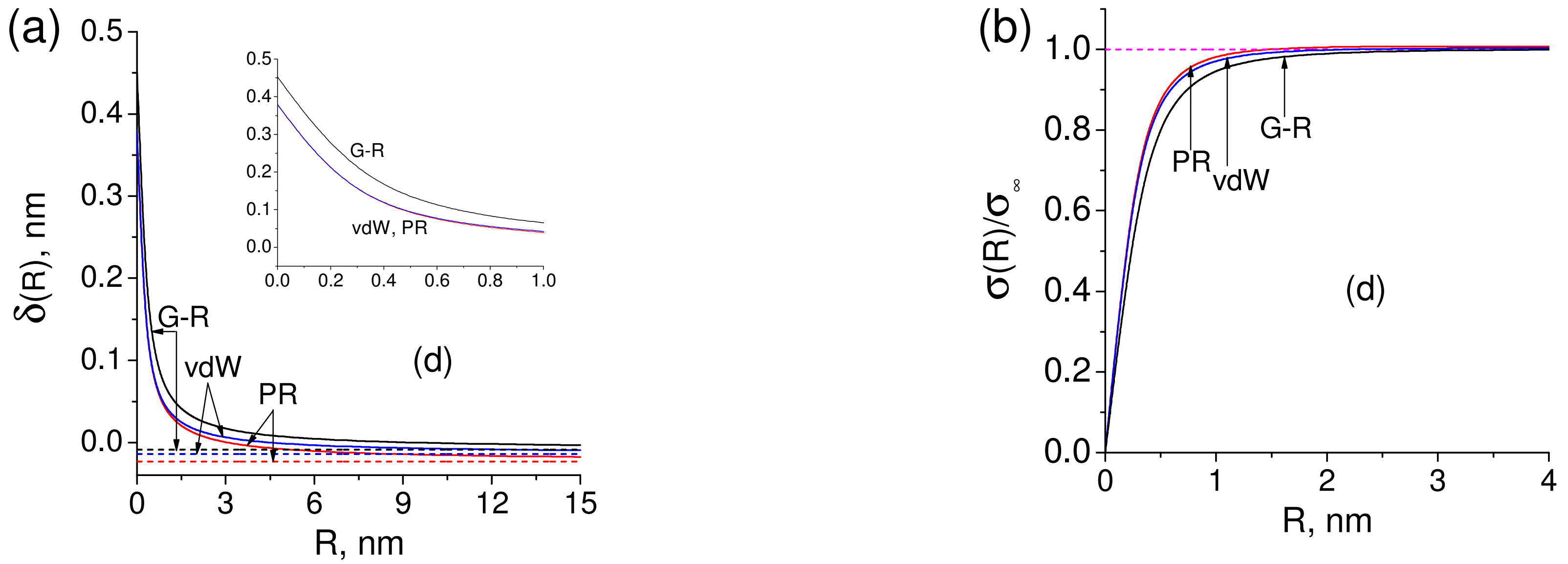

Fig. 3. Tolman lengths (a) and the corresponding dependences $\sigma(R)$ (b) in the drop case for the EOS under consideration.

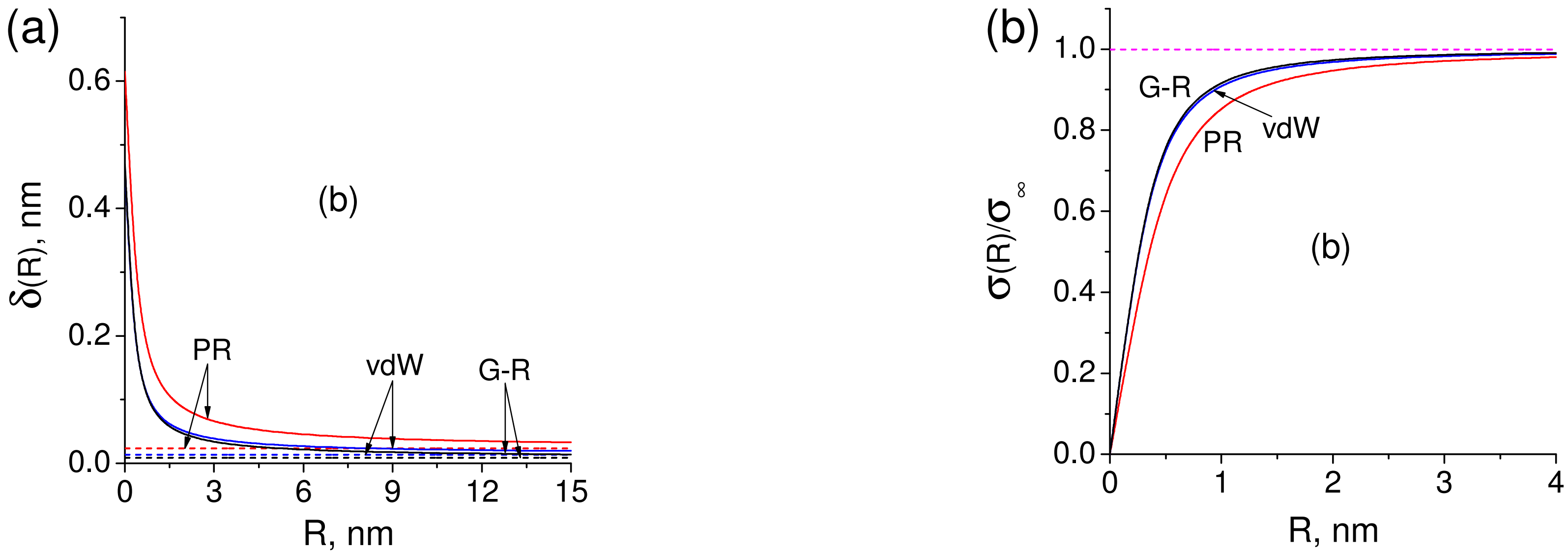

Fig. 4. Tolman lengths (a) and the corresponding dependences $\sigma(R)$ (b) for bubbles.

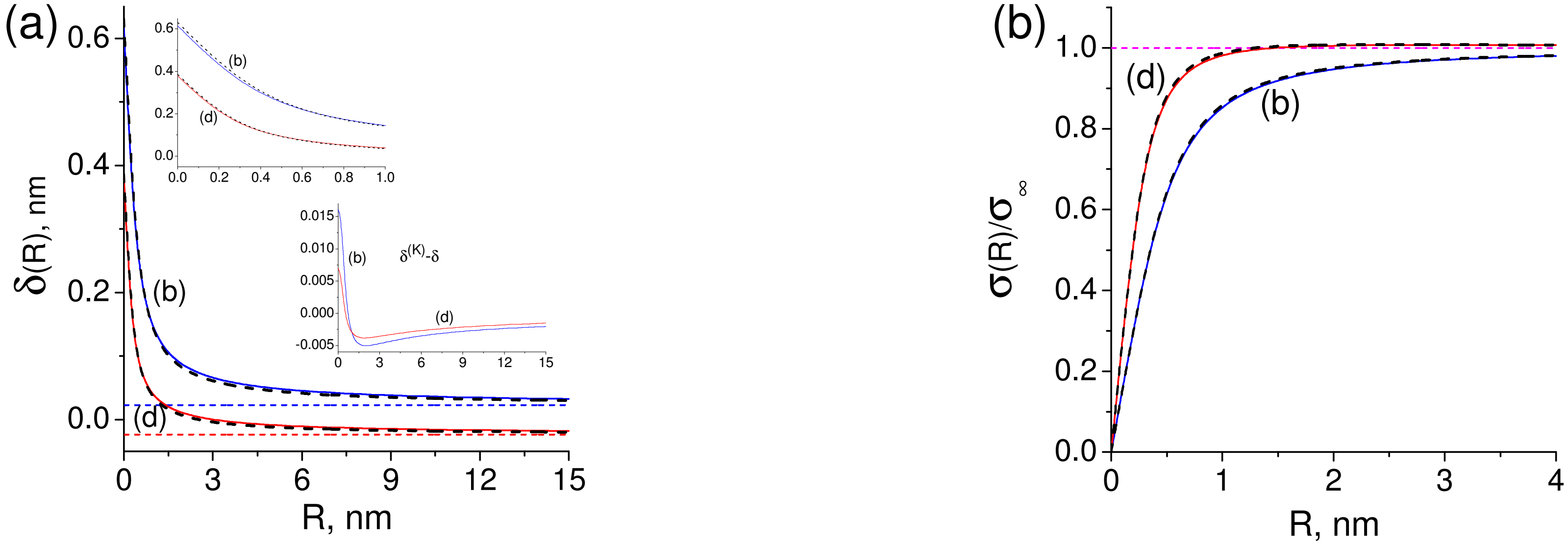

Fig. 5. (a) Exact (solid) and approximate, Eq. (68), (heavy dashed) Tolman lengths, $\delta(R)$ and $\delta^{(K)}(R)$, respectively, for the drop and bubble cases and PR EOS; the difference $\delta^{(K)}(R)-\delta(R)$ is shown in the inset. (b) Corresponding dependences $\sigma(R)$.

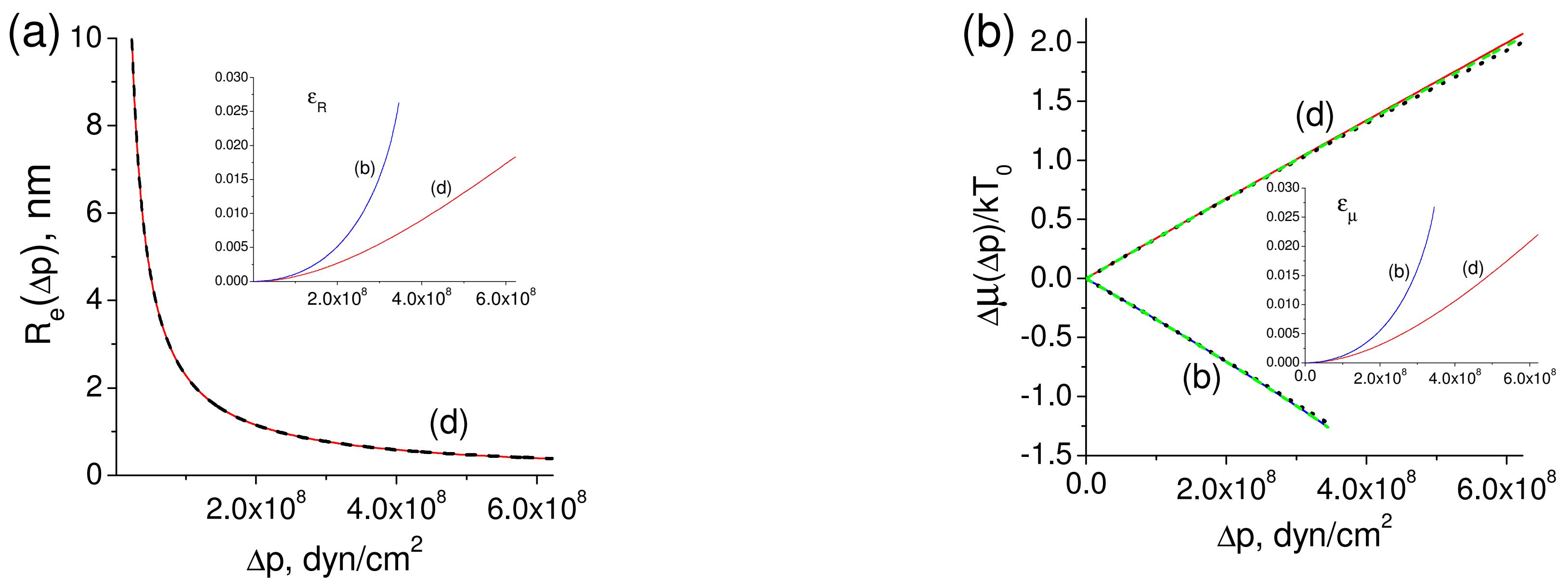

Fig. 6. (a) Exact (solid) and approximate, Eq. (65), (dashed) dependences $R_e(\Delta p)$ for the drop case and PR EOS; their relative difference $\varepsilon_R(\Delta p)$ for drops and bubbles is shown in the inset. (b) Functions $\Delta\mu(\Delta p)$ (solid), $\Delta\mu_-(\Delta p)$ (dashed), and $\Delta\mu_{ap}(\Delta p)$ (dotted) for drops and bubbles; the relative difference $\varepsilon_\mu$ of $\Delta\mu_-$ and $\Delta\mu_{ap}$ is shown in the inset.

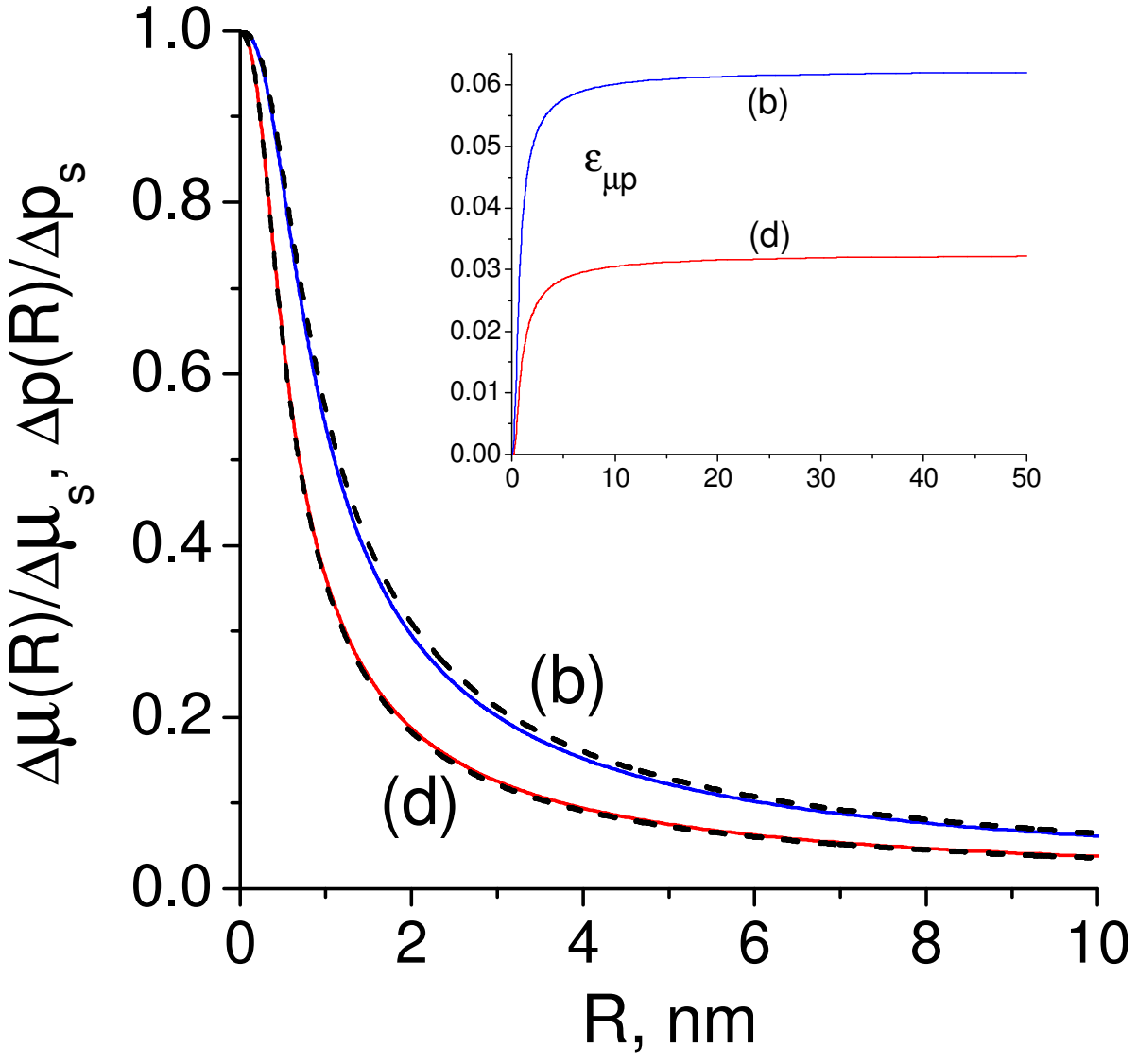

Fig. 7. Reduced chemical potential difference (solid), reduced Laplace pressure (dashed) and their relative difference (inset) for drops and bubbles.

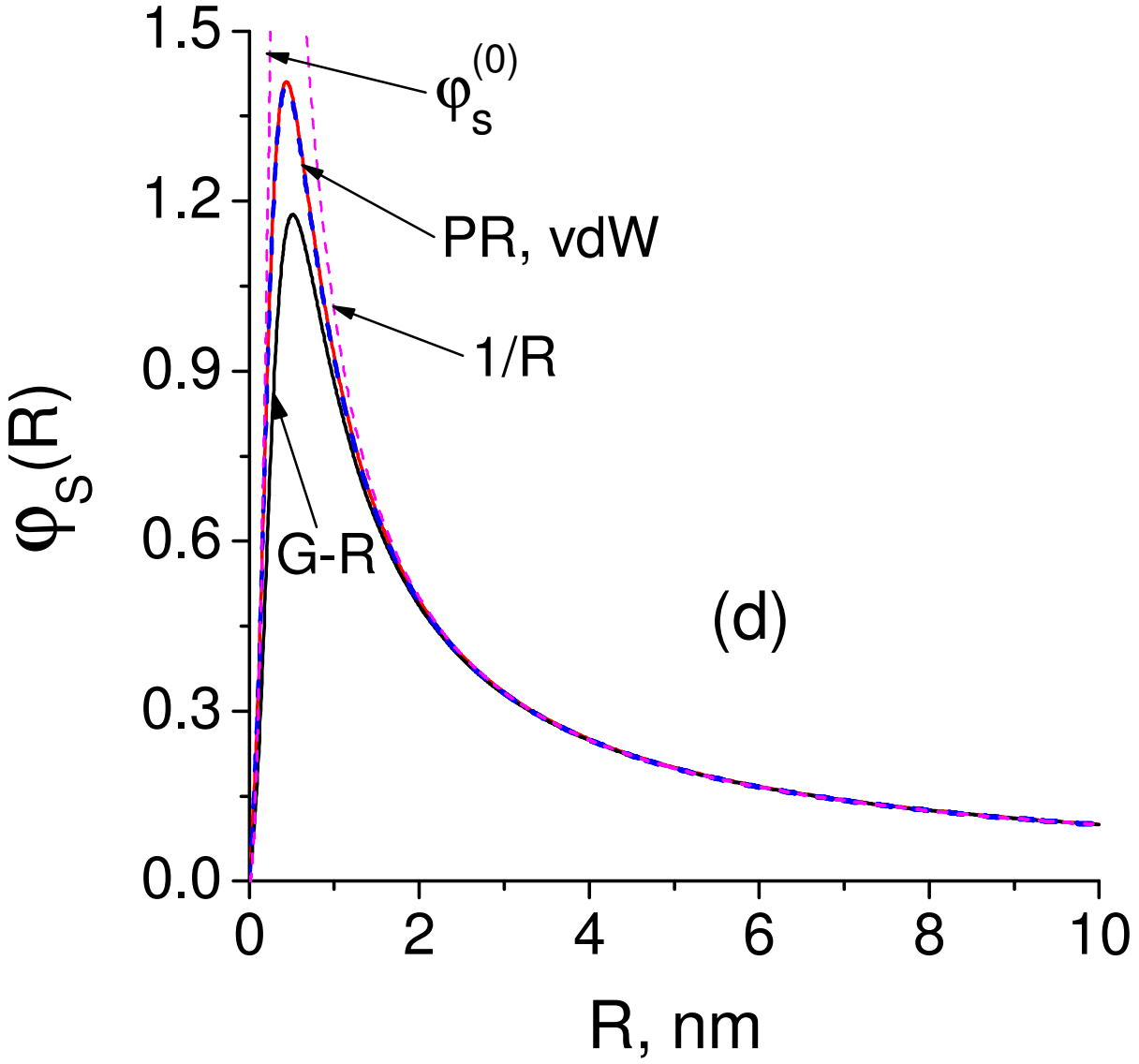

Fig. 8. Function $\varphi_S(R)$ for various EOS and its asymptotics at small and large radii.